\documentclass[aps,prl,reprint,superscriptaddress,amsmath,amsfonts,showpacs,floatfix]{revtex4-1}
\usepackage{graphicx}

\newcommand{\ket}[1]{\left| #1 \right\rangle} 
\DeclareMathOperator{\arcsec}{arcsec} 
\DeclareMathOperator{\im}{\mathrm{i}} 
\DeclareMathOperator{\perm}{Perm} 
\DeclareMathOperator{\arccot}{arccot} 
\newcommand{\abs}[1]{\left\lvert {#1} \right\rvert} 
\newcommand{\opa}{\hat{a}} 
\newcommand{\opad}{\hat{a}^\dagger} 
\newcommand{\op}[1]{\hat{#1}} 
\newcommand{\opd}[1]{\hat{#1}^\dagger} 

\def\opex{ Opt.\ Express }

\def\ol{ Opt.\ Lett.\ }

\def\pra{ Phys.\ Rev.\ A }
\def\prb{ Phys.\ Rev.\ B }

\def\prl{ Phys.\ Rev.\ Lett.\ }

\begin{document}

\title{Complete three photon Hong-Ou-Mandel interference at a three port device}

\author{S.~M\"ahrlein}
\affiliation{Institut f\"{u}r Optik, Information und Photonik, Universit\"{a}t Erlangen-N\"{u}rnberg, 91058 Erlangen, Germany}
\affiliation{Erlangen Graduate School in Advanced Optical Technologies (SAOT), Universit\"at Erlangen-N\"urnberg, 91052 Erlangen, Germany}

\author{J.~von~Zanthier}
\affiliation{Institut f\"{u}r Optik, Information und Photonik, Universit\"{a}t Erlangen-N\"{u}rnberg, 91058 Erlangen, Germany}
\affiliation{Erlangen Graduate School in Advanced Optical Technologies (SAOT), Universit\"at Erlangen-N\"urnberg, 91052 Erlangen, Germany}

\author{G.~S.~Agarwal}
\affiliation{Department of Physics, Oklahoma State University, Stillwater, Oklahoma 74078, USA}

\date{\today}


\begin{abstract}
We report the possibility of completely destructive interference of three indistinguishable photons on a three port device providing a generalisation of the well known Hong-Ou-Mandel interference of two indistinguishable photons on a two port device. Our analysis is based on the underlying mathematical framework of SU(3) transformations rather than SU(2) transformations. We show the completely destructive three photon interference for a large range of parameters of the three port device and point out the physical origin of such interference in terms of the contributions from different quantum paths. As each output port can deliver zero to three photons the device generates higher dimensional entanglement. In particular, different forms of entangled states of qudits can be generated depending on the device parameters. Our system is different from a symmetric three port beam splitter which does not exhibit a three photon Hong-Ou-Mandel interference.
\end{abstract}

\maketitle

\section{Introduction}
The Hong-Ou-Mandel (HOM) effect \cite{Hong1987,Shih1988}, i.e., the completely destructive interference of two independent but indistinguishable photons, brought a paradigm shift to the field of quantum optics. Until the demonstration of the HOM effect the interference of independent photons was considered to be impossible. Such an interference effect manifests itself in the study of photon correlations rather than in intensity measurements. More specifically, if two single photons are sent from two different ports of a 50/50 beam splitter then the number of coincidence events at the two output ports vanishes. This follows from the fact that if two photons are indistinguishable with respect to their wavelength and polarisation and their wave packages overlap in time then the two different quantum paths interfere so that the two photons will never leave the beam splitter at different ports. If one of these parameters is changed the photons become distinguishable and the dip in the observed coincidence rate starts to disappear.

The effect is quite versatile and has been observed in a very wide class of systems. Besides systems with discrete optical elements like beam splitters it has been studied in other optical setups such as in integrated devices like evanescently coupled waveguides \cite{Politi2008,Rai2008,Bromberg2009} and coupled plasmonic systems \cite{Tame2013,Gupta2014,Martino2014,Fakonas2014}. It has also been studied in the radiation from two trapped ions \cite{Maunz2007}, atoms \cite{Gillet2010,Wiegner2010,Hofmann2012} , quantum dots \cite{Patel2010,Gold2014} and for two different kinds of sources \cite{Li2008,Laiho2009,Wiegner2010a}.

Since the original work of HOM one also has examined the kind of interference that can take place if two single photons are replaced by say two photons on each port or say by one at one port and two at the other port. Hereby one has found very interesting quantum interference effects depending on the beam splitter reflectivity \cite{Ou1999,Wang2005a,Liu2007}. Another interesting possibility occurs if $n$ photons arrive at each port of a 50/50 beam splitter - in this case the output ports never have odd numbers of photons \cite{Agarwal2012}.

In this letter we report a three photon interference effect which is in the original spirit of the HOM effect - we examine the completely destructive interference of three indistinguishable photons on a three port device. We thus shift the focus from a two port device to a three port device. This brings a key change to the underlying mathematical framework as we work with SU(3) transformations rather than SU(2) transformations. We specifically examine a three port integrated device consisting of a small array of three single mode evanescently coupled waveguides as these are relatively easy to fabricate \cite{Suzuki2006,Tanzilli2012}. Although we tailor our discussion to coupled waveguide systems the results will be applicable to a wide class of bosonic systems described by the Hamiltonian (\ref{eq:Hint}). For the three port device we have found an analytical expression for the completely destructive three photon interference. Thereby we produce a variety of two and three photon entangled states at the output ports.

Our three port network is different from the symmetric multiport beam splitter which has been extensively studied for HOM like interferences \cite{Lim2005a,Tichy2010}, as for the three port system such a splitter does not exhibit a perfect three photon HOM interference. On the other hand, Campos \cite{Campos2000} using the idea of Reck et al. \cite{Reck1994} constructed a SU(3) transformation involving beam splitters and phase shifters which leads to three photon HOM interference. Further, Tan et al. \cite{Tan2013} showed how the SU(3) transformation involving beam splitters and phase shifters can lead to perfect photon interference depending on specific values of the parameters of SU(2) transformations. While the papers of Campos \cite{Campos2000} and Tan et al. \cite{Tan2013} concentrate on configurations using beam splitters we investigate integrated devices. Such integrated devices are now used by many experimental groups. Studies of multi photon interferences in integrated devices have been reported, e.g., in \cite{Weihs1996,Politi2008,Bromberg2009,Peruzzo2011,Meany2012,Broome2013,Spring2013,Tillmann2013,Crespi2013,Chaboyer2014}. Note that the examination of multiport systems is also important in the context of Bosonic sampling \cite{Broome2013,Spring2013,Tillmann2013,Crespi2013,Tamma2014} where the output distribution of Fock input states is sampled what is exponentially hard to predict with classical computations.

While our investigated setup is similar to the experiment studied in \cite{Spagnolo2013}, which relies on a 3D geometry in order to couple all three modes to each other, we will use a more simple 2D structure, where the outer modes are coupled to the inner mode but not to each other. With the setup of \cite{Spagnolo2013} it is possible to suppress \cite{Tichy2010} states of the form $\ket{2,1,0}$, which contain two, one and zero photons in the different output modes, but the output state will still contain a $\ket{1,1,1}$-term corresponding to the coincidence detection of all three photon in the three different output modes. We will show that for a whole range of parameters of the 2D waveguide structure our system can suppress this coincidence event which corresponds to the original Hong-Ou-Mandel effect \cite{Hong1987,Shih1988} extended to three interfering photons.

\section{Model and time evolution}

\begin{figure}
	\centering
	\includegraphics{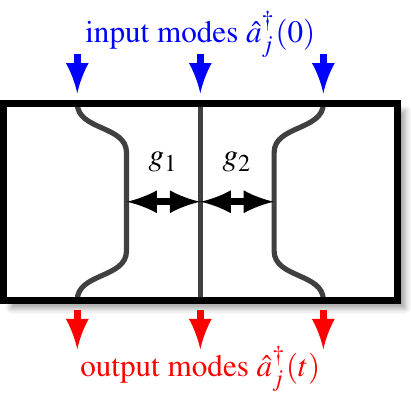}
	\caption{Scheme of a $3 \times 3$ waveguide array with continuous coupling between the inner mode and the outer modes}
	\label{fig:waveguide}
\end{figure}

The investigated system consists of a 2D $3 \times 3$ waveguide array (three input modes and three output modes) with continuous evanescent coupling between the inner mode and the outer modes (see Fig. \ref{fig:waveguide}). We assume identical single mode waveguides with uniform coupling throughout the waveguide array. We also assume that the waveguide mode's frequency is matched to the frequency of the input field. The coupling strength is given by the coupling coefficients $g_1$ (between the first and second mode) and $g_2$ (between the second and third mode) and is basically determined by the distance between the guides. The calculation of the coupling parameters is text book material and can be found, e.g., in \cite{Saleh2007}. Note that the coupling between the two outer modes is negligible for our geometry.

The interaction Hamiltonian for this system reads
\begin{equation}
	\op{H}_\text{int} = \hbar \left( g_1 \opa_1 \opad_2 + g_2 \opa_2 \opad_3 + g_1^* \opa_2 \opad_1  + g_2^* \opa_3 \opad_2 \right).
\label{eq:Hint}
\end{equation}\\
Each part of Eq. (\ref{eq:Hint}) stands for the annihilation of a photon in a certain mode and the creation of a photon in a neighbouring mode associated with the corresponding coupling strength $g_{1/2}$. Although $g_{1}$ and $g_{2}$ are real for waveguide systems we keep these parameters complex, since SU(3) Hamiltonians occur for many physical systems - for example $N$ identical three level atoms interacting with external fields of different phases. We further add that the Hamiltonian that we use adequately describes the investigated waveguide system although it is not the most general SU(3) Hamiltonian. The calculations can be done for the most general case, however, we will see that the Hamiltonian in Eq. (\ref{eq:Hint}) already leads to a number of very interesting interference effects by keeping the analysis more simple.

In order to analyse the time dependent evolution of the system we switch to the Heisenberg picture. To simplify the calculation we define a vector $\vec{a} = \left( \opad_1, \, \opad_2, \, \opad_3 \right)^T$ so that the time evolution is governed by
\begin{equation}
	\frac{\text{d}}{\text{d}t} \vec{a}(t) = \im \underbrace{\left(\begin{matrix} 0 & g_1 & 0 \\ g^*_1 & 0 & g_2 \\ 0 & g^*_2 & 0 \end{matrix} \right)}_{M} \vec{a}(t),
	\label{eq:Heisenberg2}
\end{equation}
where  the interaction time $t$ is determined by the length of the waveguide. The equation can easily be solved using the exponential ansatz $V = e^{-\im M t}$, which allows to rewrite the creation operators $\opad_j(0)$ at time $t=0$ in terms of creation operators $\opad_j(t)$ at time $t$ via $\vec{a}(0) = V\cdot \vec{a}(t)$. The solution of this equation yields the explicit form of the matrix $V$
\begin{widetext}
\begin{equation}
	V = \left(\begin{matrix}  \cos^2 (\theta) \cos (G) + \sin^2 (\theta) &  -\im \cos (\theta) \sin (G) \, e^{\im \psi} & \cos (\theta) \sin (\theta) e^{\im (\varphi + \psi)} \left( \cos (G) - 1 \right) \\  -\im \cos (\theta) \sin (G) e^{-\im \psi} &  \cos (G) &  -\im \sin (\theta) \sin (G) e^{\im \varphi} \\   \cos (\theta) \sin (\theta) e^{-\im (\varphi + \psi)} \left( \cos (G) - 1 \right) & -\im \sin (\theta) \sin (G) e^{-\im \varphi} &  \sin^2 (\theta) \cos (G) + \cos^2 (\theta) \end{matrix} \right).
	\label{eq:Uexpl}
\end{equation}
\end{widetext}
where $g_1 \cdot t = G \, \cos (\theta) \, e^{\im \psi}$ and $g_2 \cdot t = G \, \sin (\theta) \, e^{\im \varphi}$. From this expression it is easy to see  that the evolution given by $V$ is periodic with respect to $G$ and $\theta$. The periodicity in $\theta$ arises due to the change to polar coordinates while the periodicity in $G$ is linked to the interpretation of $G$ as overall scaling factor, i.e., for example to variations of the interaction lengths of the waveguide array. Specifically, for $G=\sqrt{\left|g_1\right|^2+\left|g_2\right|^2} \cdot t=2\pi \cdot m$, with $m \in \mathbb{N}$, we get $\sin(G)=0$ and $\cos(G)=1$, so that $V$ reduces to the identity matrix. This means that the system returns to its initial state, what is also known as self-imaging or revivals of multi mode interferences \cite{Heaton1999, Poem2011}. Note that, as expected, the transformation $V$ is unitary - this is because we are considering a lossless device.

\begin{figure}
	\centering
	\includegraphics[width=0.45\textwidth]{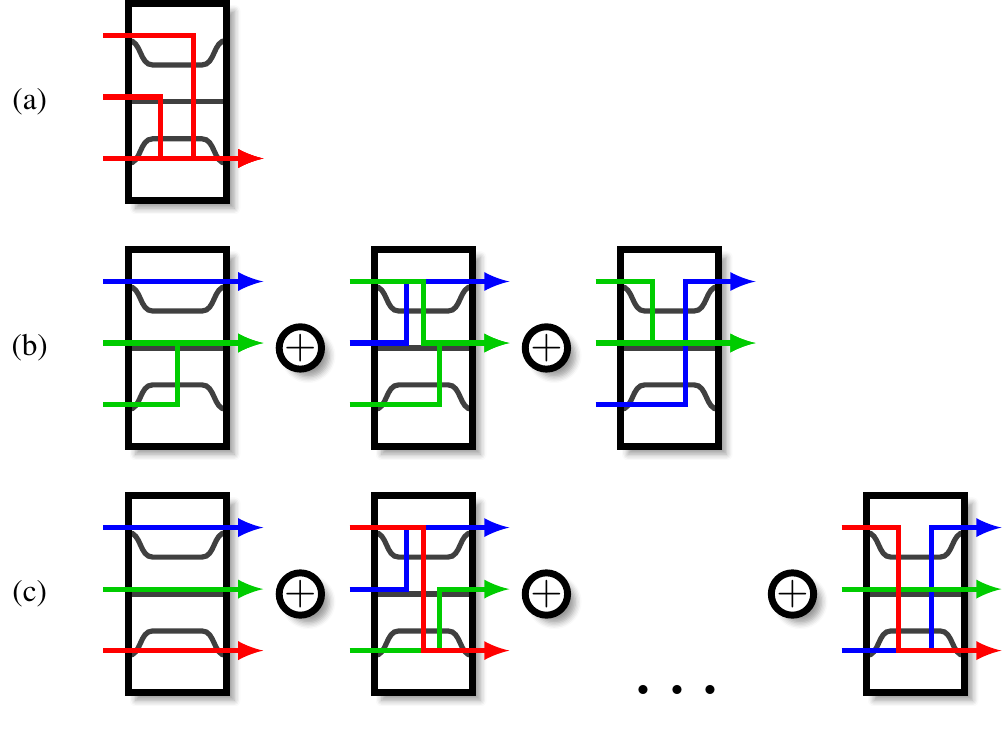}
	\caption{Three examples for different indistinguishable quantum paths leading to the same output state $\ket{300}$ in (a), $\ket{021}$ in (b) and $\ket{111}$ in (c)}
	\label{fig:qpaths}
\end{figure}

\section{Three single photon input}
Next we focus on an input state where at each input port a single photon is coupled into the waveguide at $t=0$. The wave function of this state is given by	$\ket{\psi_\text{in}} = \opad_1 (0) \opad_2 (0) \opad_3 (0) \ket{0}$. Via the transformation matrix $V$ we can easily calculate the general form of the output state
\begin{equation}
	\ket{\psi_\text{out}} =\sum_{l=1}^{3} V_{1l} \opd{a}_l(t) \cdot \sum_{m=1}^{3} V_{2m} \opd{a}_m(t) \cdot \sum_{n=1}^{3} V_{3n} \opd{a}_n(t) \ket{0},
	\label{eq:Ouput}
\end{equation}
where $V_{mn}$ is the matrix element in the $m$th row and the $n$th column of $V$ in Eq.~(\ref{eq:Uexpl}). As the transformation matrix $V$ only acts on the creation operators of the three different modes  in Eq.~(\ref{eq:Ouput}) a vacuum state is not transformed by the time evolution governed by the Hamiltonian of Eq.~(\ref{eq:Hint}). The general output state is a superposition of all possible distributions of the three photons among the three output modes with respective coefficients depending on the explicit form of $V$. The general form of the output state reads
\begin{equation}
\begin{split}
	\ket{\psi_\text{out}} = &c_{300} \ket{3,0,0} +  c_{030} \ket{0,3,0} +\ldots \\ + &c_{210} \ket{2,1,0} + \ldots + c_{111} \ket{1,1,1},
	\label{eq:GenOuput}
\end{split}
\end{equation}
where corresponding coefficients can be calculated by explicitly expanding Eq. (\ref{eq:Ouput}) or alternatively using a formalism for linear optical networks involving permanents \cite{Scheel2004}.

A permanent of a matrix is equal to its determinant but without the sign of the permutation taken into account. For a $N\times N$ matrix $A$ it is given by
\begin{equation}
	\perm A = \sum_{\sigma} \prod_{j=1}^N A_{j\sigma(j)},
	\label{eq:permanent}
\end{equation}
where the sum runs over all possible permutations $\sigma$ of the set $\{1,2,\ldots,N \}$.

The coefficients $c_{klm}$ of Eq. (\ref{eq:GenOuput}) can be expressed by permanents of matrices $V^{\{k,l,m\}}$, where $k$, $l$ and $m$ are the number of photons in the three output modes so that $k+l+m=3$. Hereby $V^{\{k,l,m\}}$ is a $3 \times 3$ matrix and is constructed via the transformation matrix of Eq. (\ref{eq:Uexpl}). It consists of $k$ copies of the first column of $V$, $l$ copies of the second column of $V$ and $m$ copies of the third column of $V$ \cite{Scheel2004}. Dividing the permanent of $V^{\{k,l,m\}}$ by a normalisation factor yields the final expression for the coefficients
\begin{equation}
	c_{klm}= \frac{\perm V^{\{k,l,m\}}}{\sqrt{k!l!m!}}.
	\label{eq:c_klm}
\end{equation}
One can show that the absolute value of these coefficients depends only on $G$ and $\theta$ but not on the phases  $\psi$ and $\varphi$ of the coupling coefficients $g_{1/2}$ which will only have an impact on the phases of the coefficients $c_{klm}$.

Note that Eq. (\ref{eq:c_klm}) only holds true for the input state $\ket{1,1,1}$. However, as shown in \cite{Scheel2004}, the formalism involving permanents can be expanded to arbitrary initial states by additional consideration of the rows of the transformation matrix corresponding to the input state.

The permanents can be understood as produced by the coherent superposition of indistinguishable quantum paths (excluding the normalization factor) leading to the same output state. To illustrate this we consider in the following three examples, starting with the coefficient $c_{300}$.  To calculate this coefficient, associated with the state $\ket{300}$, one first has to construct the matrix $V^{\{300\}}$ containing three copies of the first row of the transformation matrix $V$. It therefore reads
\begin{equation}
	V^{\{3,0,0\}} = \left(\begin{matrix} V_{11} & V_{11} & V_{11} \\ V_{21} & V_{21} & V_{21} \\ V_{31} & V_{31} & V_{31} \end{matrix} \right),
	\label{eq:V300}
\end{equation}
where $V_{nm}$ are the corresponding matrix elements of $V$ in Eq. (\ref{eq:Uexpl}) determined by the transition amplitude for a photon initially in mode $n$ to exit in mode $m$. From Eq.~(\ref{eq:V300}) the permanent of $V^{\{300\}}$ can be calculated yielding
\begin{equation}
	\perm V^{\{3,0,0\}} = 6 V_{11} V_{21} V_{31}.
	\label{eq:perm300}
\end{equation}
This expression corresponds to the only possible quantum path where a photon in the first mode stays in the first mode and a photon of the second mode and a photon of the third mode also exits the waveguide array in the first mode as illustrated in Fig. \ref{fig:qpaths}(a). As only a single quantum path appears for this coefficient no interference is observed in this case.

The calculation of any other coefficient like, e.g., $c_{021}$ for the state $\ket{0,2,1}$, follows the same structure: First one constructs the matrix $V^{\{0,2,1\}}$ by taking two copies of the second row and one copy of the third row of $V$
\begin{equation}
	V^{\{0,2,1\}} = \left(\begin{matrix} V_{12} & V_{12} & V_{13} \\ V_{22} & V_{22} & V_{23} \\ V_{32} & V_{32} & V_{33} \end{matrix} \right)
	\label{eq:V021}
\end{equation}
and then calculates its permanent, yielding in this case
\begin{equation}
\begin{split}
	\perm V^{\{0,2,1\}} = 2 \big(  &V_{12} V_{22} V_{33} + V_{12} V_{23} V_{32}  \\  + &V_{13} V_{22} V_{32} \big).
	\label{eq:perm021}
\end{split}
\end{equation}
Now three different indistinguishable quantum paths appear which have to be added coherently as shown in Fig. \ref{fig:qpaths}(b), what leads to interference effects. The three quantum paths correspond to the three possibilities how three photons entering the waveguide array initially in three different modes  exit the array in the output state $\ket{021}$.

\begin{figure}
	\centering
		\includegraphics{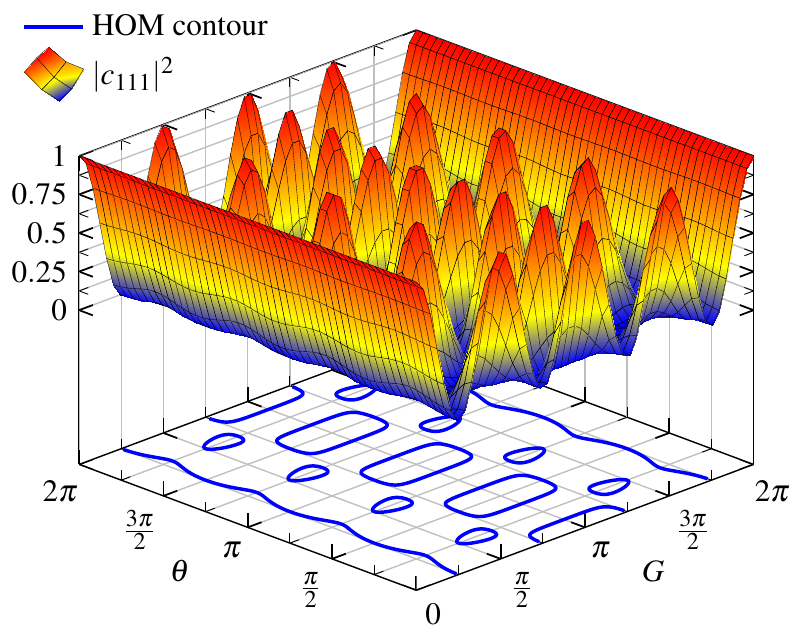}
	\caption{Plot of the three photon coincidence probability $\abs{c_{111}}^2$ and the HOM contour, where the condition $\abs{c_{111}}^2$=0 is fulfilled.}
	\label{fig:HOMcontour}
\end{figure}

As a last example we calculate the coefficient $c_{111}$ corresponding to a three mode coincidence event. As the matrix $V^{\{111\}}$ consists of a copy of each row it is equal to $V$ and its permanent is calculated to be
\begin{equation}
\begin{split}
	\perm V^{\{1,1,1\}} = &V_{11}V_{22}V_{33}+V_{12}V_{23}V_{31}+V_{13}V_{21}V_{32}  \\  + &V_{11}V_{23}V_{32}+V_{12}V_{21}V_{33}+V_{13}V_{22}V_{31} \, .
\end{split}
\label{eq:perm111}
\end{equation}
It consists of six different terms each corresponding to a different indistinguishable quantum path leading to the same final state $\ket{1,1,1}$. For example, the first term corresponds to the case where all three photons exit the waveguide in the same mode they came in, the second term corresponds to the case where the photon in the first/second/third mode switches to the second/third/first mode etc., as depicted in Fig. \ref{fig:qpaths}(c). As in the second example the multiple indistinguishable quantum paths can interfere with each other. In particular, they can interfere in a completely destructive way. This configuration will be analyzed in the next section.

\section{Three photon Hong-Ou-Mandel interference}
In the following we investigate the particular situation displaying a three photon Hong-Ou-Mandel (HOM) interference. In analogy to the original two photon Hong-Ou-Mandel experiment \cite{Hong1987,Shih1988} where two photons are never detected simultaneously at the two different output modes of a 50/50 beam splitter, the probability for all three photons leaving the waveguide at the three output ports vanishes if
\begin{equation}
	c_{111} \overset{!}{=} 0.
	\label{eq:c111HOM}
\end{equation}
 To analyse the conditions for the three photon HOM interference, we have to calculate $c_{111}$ explicitly. From Eqs.~(\ref{eq:c_klm}) and (\ref{eq:perm111}) we find
\begin{equation}
\begin{split}
	c_{111} = &V_{11}V_{22}V_{33}+V_{12}V_{23}V_{31}+V_{13}V_{21}V_{32}  \\  + &V_{11}V_{23}V_{32}+V_{12}V_{21}V_{33}+V_{13}V_{22}V_{31}.
\end{split}
\label{eq:c111}
\end{equation}
By inserting the expression for the various matrix elements $V_{mn}$ of Eq. (\ref{eq:Uexpl}) and solving Eq. (\ref{eq:c111HOM}) we can find an analytical expression for the HOM contour in the variable space ($G$, $\theta$), where all states $\ket{\psi_\text{out}(G,\theta)}$ have a vanishing $c_{111}$ coefficient:
\begin{widetext}
\begin{equation}
	{\textstyle \theta (G) = n\pi \pm \arcsec \left[ 4\bigg/\sqrt{8 \pm \frac{\sqrt{2} \csc ^4\left(\frac{G}{2}\right) \sqrt{\sin ^4\left(\frac{G}{2}\right) (20 \cos (G)+3 (8 \cos (2 G)+4 \cos (3 G)+3 \cos (4 G)+5))}}{3 \cos (G)+2}}\right]}
	\label{eq:thetaG}
\end{equation}
\end{widetext}

As can be seen from Eq. (\ref{eq:thetaG}) completely destructive three photon HOM interference can take place for a large range of the parameters $g_1$ and $g_2$. Note that for some values of $G$ these equations would result in a complex valued $\theta$ and are therefore not considered as a solution. Fig. \ref{fig:HOMcontour} shows a plot of the HOM contour, displaying in particular the periodicity in $G$ and $\theta$ discussed in Sect.~2.

\section{Interesting states on the Hong-Ou-Mandel contour}
Finally we investigate the states determined by the HOM contour. In the original HOM experiment a maximally entangled state of the form $\propto \ket{2,0} - \ket{0,2}$ is produced at the output \cite{Hong1987,Shih1988}. Similar states can be found in the case of a three photon interference. Additionally to the condition $c_{111}=0$ we find that at certain points some coefficients $c_{klm}$ of Eq. (\ref{eq:GenOuput}) will vanish as well, so that further terms are suppressed. Other coefficients will have the same absolute value, so that the states can be written in a compact form. We found three different kinds of states fulfilling this condition, which display entanglement between two and possibly three output modes.

\begin{figure*}
	\centering
		\includegraphics{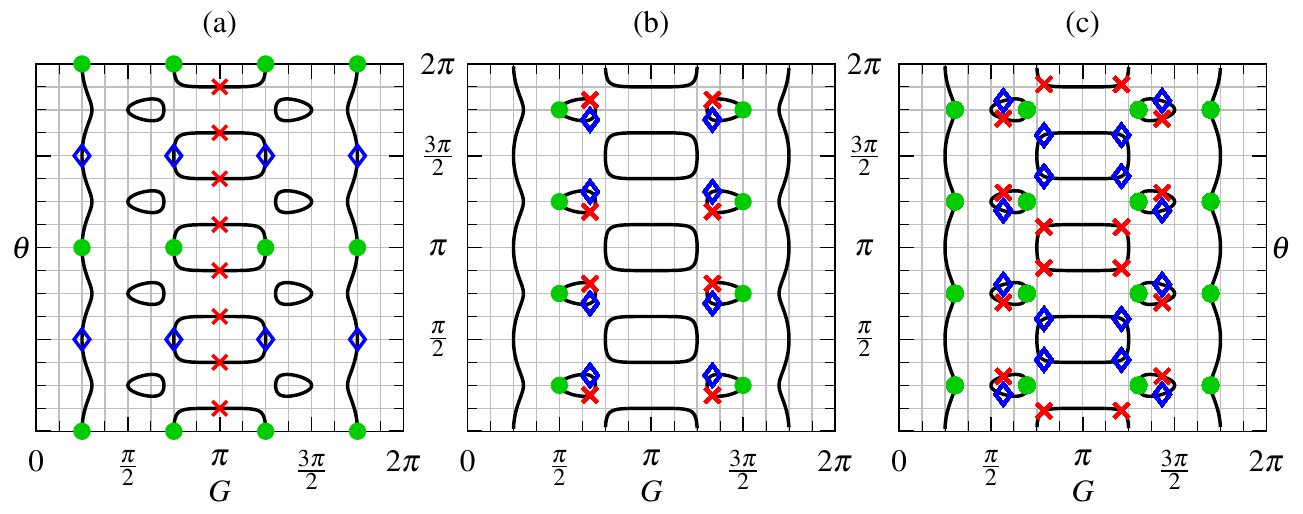}
	\caption{Plots of the HOM contour and the coordinates of the investigated states (red crosses, green dots and blue diamonds), which display bipartite entanglement (a) or possibly tripartite entanglement (b)/(c).}
	\label{fig:sysent}
\end{figure*}

In Fig.~\ref{fig:sysent}(a), where the HOM contour is shown, some coordinates are marked by points where one can find maximally bipartite entangled states. A closer investigation yields that for all these  coordinates we find states of the form 
\begin{equation}
	\ket{\psi_\text{out}} = \frac{1}{\sqrt{2}} \Bigl( \ket{2_j,0_k} + \ket{0_j,2_k} \Bigr) \ket{1_l}
\label{eq:ent2}
\end{equation}
with $j=1$, $k=3$ and $l=2$ at the red crosses, with $j=1$, $k=2$ and $l=3$ at the green dots and with $j=2$, $k=3$ and $l=1$ at the blue diamonds. Note that we neglected the global and relative phase factors of each state, as we just want to focus on the structure of the states (see the appendix for the exact analytical expressions for each coefficient and the corresponding coordinates). All the three states have a similar form where one mode containing one photon is separable while the remaining two modes are in a maximally entangled state. Depending on the phase $\psi / \varphi$ of the coupling coefficients $g_{1/2}$ the relative phase between the non-separable states can be varied. A closer look at the transformation matrix $V$ of Eq.~(\ref{eq:Uexpl}) reveals that in these cases the absolute values of the matrix elements are given by $\left|V_{ll}\right|=1$, $\left|V_{jk}\right|=\left|V_{kj}\right|=\left|V_{jj}\right|=\left|V_{kk}\right|=\frac{1}{\sqrt{2}}$ while all remaining matrix elements vanish. This means that the mode $l$ is decoupled from the system as the photon initially entering the waveguide array in this mode will always exit the waveguide array in the same mode. However, the modes $k$ and $j$ are mixed with a 50/50 ratio corresponding to the original two photon HOM interference effect. This can also be seen from Eq.~(\ref{eq:c111}) where under the same conditions only two of the six quantum paths survive and interfere completely destructively. For $l = 2$, the corresponding quantum paths are the two paths, one at the extreme left and the other at the extreme right, in Fig. \ref{fig:qpaths}(c). At the coordinates of the green dots (blue diamonds) it is evident that the outer mode mode $l=3$ ($l=1$) is physically decoupled since the coupling parameter $g_2$ ($g_1$) vanishes. However, at the red crosses both coupling constants $g_1$  and $g_2$ are physically present but the inner mode $l=2$ is effectively decoupled from the system and merely acts as a mediator between the outer modes. Therefore, it is more appropriate to speak of a two photon interference at these coordinates although three photons are present. The physical decoupling of modes has also been discussed in Campos \cite{Campos2000} in the context of beam splitter devices. Note that all the discussed states are created in a deterministic way so that no post selection is required.

In Figs.~\ref{fig:sysent}(b) and \ref{fig:sysent}(c) the coordinates of possible tripartite entangled states along the HOM contour are displayed. Note that we have three modes and each mode has four possible states corresponding to the occupation of 0, 1, 2 and 3 photons. We are thus dealing with higher dimensional entanglement of three qudits ($d=4$). A complete classification of the classes of entangled states for qudits ($d=4$) does not exist. However the structure of tripartite states generated for the case of Figs.~\ref{fig:sysent}(b) and \ref{fig:sysent}(c) suggests three qudit ($d=4$) entanglement. The form of the states of Fig. \ref{fig:sysent}(b) and their coordinates read
\begin{equation}
\begin{split}
	\ket{\psi_\text{out}}= \frac{\sqrt{3}}{4} &\Bigl(\ket{3_j,0_k} + \ket{0_j,3_k} \Bigr) \ket{0_l} \\
	+ \frac{1}{4} &\Bigl( \ket{2_j,1_k} + \ket{1_j,2_k} \Bigr) \ket{0_l} \\ 
	+ \frac{1}{2} &\Bigl( \ket{1_j,0_k} + \ket{0_j,1_k} \Bigr) \ket{2_l}
\end{split}
	\label{eq:ent31}
\end{equation}
with $j=2$, $k=3$ and $l=1$ at the red crosses, with $j=1$, $k=3$ and $l=2$ at the green dots and with $j=1$, $k=2$ and $l=3$ at the blue diamonds. The form of the wave function of the states of Fig. \ref{fig:sysent}(c) and their coordinates are given by
\begin{equation}
\begin{split}
	\ket{\psi_\text{out}} = \frac{1}{3\sqrt{2}} &\Bigl( 2 \ket{3_j,0_k,0_l} + \ket{0_j,3_k,0_l} + \ket{0_j,0_k,3_l} \Bigr) \\
	+ \frac{1}{\sqrt{6}} &\Bigl( \ket{1_j,0_l} + \ket{0_j,1_l} \Bigr) \ket{2_k} \\
	+ \frac{1}{\sqrt{6}} &\Bigl(\ket{1_j,0_l} + \ket{0_j,1_k} \Bigr) \ket{2_l}
\end{split}
	\label{eq:ent32}
\end{equation}
with $j=1$, $k=2$ and $l=3$ at the red crosses, with $j=2$, $k=1$ and $l=3$ at the green dots and with $j=3$, $k=1$ and $l=2$ at the blue diamonds. As before we neglected the global and relative phase factors of each state for simplicity (see the appendix for the exact analytical expressions for each coefficient and the corresponding coordinates). We have written the states in a way that a certain mode is always factored out in each term so that the entanglement between the two remaining modes is clearly visible; this suggests that the states are not separable and are good candidates for tripartite entanglement.

\section{Conclusion}
In conclusion, we investigated the dynamics of a 2D 3$\times$3 waveguide where the outer modes are coupled to the inner mode by evanescent coupling but not to each other. Beginning with three indistinguishable single photons at the three input ports we showed that for a wide range of waveguide parameters this leads to completely destructive three photon interference, i.e. for these parameters the photons will never leave the waveguide in three separate ports. This is a generalisation of the well know Hong-Ou-Mandel effect from two to three photons. Additionally, the produced output states consisting of three qudits ($d=4$) exhibit highly interesting structures displaying bipartite or possibly tripartite entanglement. Clearly the entanglement of qudits is a topic for future studies.


\newpage

\section*{Acknowledgments}
The authors gratefully acknowledge funding by the Erlangen Graduate School in Advanced Optical Technologies (SAOT) by the German Research Foundation (Deutsche Forschungsgemeinschaft DFG) in the framework of the German excellence initiative. We acknowledge support by the German Research Foundation (Deutsche Forschungsgemeinschaft DFG) and by the Friedrich-Alexander-Universit\"at Erlangen-N\"urnberg (FAU) within the funding programme Open Access Publishing. We thank H. de Guise, T. Meany, V. Tamma, M. Tichy for bringing their works to our attention. S. M. gratefully acknowledges the hospitality at the Oklahoma State University. G. S. A. thanks I. Walmsley and P. Mataloni for discussions on HOM interference in integrated devices.

\section*{Appendix}
For certain waveguide parameters $G$ and $\theta$ we found three interesting sets of states on the HOM contour for which the coincident event is suppressed and therefore $c_{111}$ vanishes. However, only the general form of these states was discussed but not the explicit values of the coefficients. Here we present their analytical values: Tables \ref{tab:ent2-1} to \ref{tab:ent32-3} contain the exact analytical expressions for the coefficients of Eq. (\ref{eq:ent2}) - (\ref{eq:ent32}). Note that in Eqs. (\ref{eq:ent31}) and (\ref{eq:ent32}) some signs depend on the value of $n \, \text{mod} 4$. Therefore in the second part of Table \ref{tab:ent31} and in Table \ref{tab:ent32-2} $n$ is replaced by $4\tilde{n}+0$, $4\tilde{n}+1$, $4\tilde{n}+2$ or $4\tilde{n}+3$, in which case all possible cases are considered.

\newpage

\begin{table*}
	\centering
	\footnotesize
		\begin{tabular}[b]{c|c|c|c|c}
		$G$ & $\pi \left( 2m+1 \right)$ & $\frac{\pi}{4} \left( 2m+1 \right)$ & $\frac{\pi}{4} \left( 2m+1 \right)$ \\[3pt]
		$\theta$ & $\frac{\pi}{8} \left( 2n+1 \right)$ & $\pi n$ & $\frac{\pi}{2} \left( 2n+1 \right)$\\[3pt]
		\hline
		$c_{210}$ & $\frac{(-1)^{n+1}}{\sqrt{2}} e^{-\im (\psi + \varphi)}$ & 0 & 0 \\[3pt]
		$c_{012}$ & $\frac{(-1)^{n}}{\sqrt{2}} e^{\im (\psi + \varphi)}$ & 0 & 0\\[3pt]
		$c_{201}$ & 0 & $\frac{\im (-1)^{n+m}}{\sqrt{2}} e^{-\im \psi}$ & 0 \\[3pt]
		$c_{021}$ & 0 & $\frac{\im (-1)^{n+m}}{\sqrt{2}} e^{\im \psi}$ & 0\\[3pt]
		$c_{120}$ & 0 & 0 & $\frac{\im (-1)^{n+m+1}}{\sqrt{2}} e^{-\im \varphi}$ \\[3pt]
		$c_{102}$ & 0 & 0 & $\frac{\im (-1)^{n+m+1}}{\sqrt{2}} e^{\im \varphi}$\\[3pt]
		\end{tabular}
		\caption{Table of coefficients for equation (\ref{eq:ent2}). Coefficients, which are not listed, are equal to 0.}
		\label{tab:ent2-1}
\end{table*}

\begin{table*}
	\centering
	\footnotesize
		\begin{tabular}[b]{c|c|c|c|c}
		$G$ & $\pi \left( 2m+1-\frac{1}{3}\right)$ & $\pi \left( 2m+1+\frac{1}{3}\right)$ & $\pi \left( 2m+1-\frac{1}{3}\right)$ & $\pi \left( 2m+1+\frac{1}{3}\right)$ \\[3pt]
		$\theta$ & \multicolumn{2}{c|}{$\pi n +  \arccot\left( \sqrt{2}\right)$} & \multicolumn{2}{c}{$\pi n -  \arccot\left( \sqrt{2}\right)$} \\[3pt]
		\hline
		$c_{030}$ & $\frac{\sqrt{3}}{4} e^{\im (\psi-\varphi)}$ & $\frac{\sqrt{3}}{4} e^{\im (\psi-\varphi)}$ & $-\frac{\sqrt{3}}{4} e^{\im (\psi-\varphi)}$ & -$\frac{\sqrt{3}}{4} e^{\im (\psi-\varphi)}$ \\[2pt]
		$c_{003}$ & $\im \frac{(-1)^n \sqrt{3}}{4} e^{\im (\psi+2\varphi)}$ & $-\im \frac{(-1)^n\sqrt{3}}{4} e^{\im (\psi+2\varphi)}$ & $\im  \frac{(-1)^n \sqrt{3}}{4} e^{\im (\psi+2\varphi)}$ & $-\im \frac{(-1)^n \sqrt{3}}{4} e^{\im (\psi+2\varphi)}$ \\[3pt]
		$c_{210}$ & $\frac{1}{2} e^{-\im(\psi+\varphi)}$ & $\frac{1}{2} e^{-\im(\psi+\varphi)}$ & $-\frac{1}{2} e^{-\im(\psi+\varphi)}$ & $-\frac{1}{2} e^{-\im(\psi+\varphi)}$ \\[3pt]
		$c_{201}$ & $-\im \frac{(-1)^n}{2} e^{-\im \psi}$ & $\im \frac{(-1)^n}{2} e^{-\im \psi}$ & $-\im \frac{(-1)^n}{2} e^{-\im \psi}$ & $\im \frac{(-1)^n}{2} e^{-\im \psi}$ \\[3pt]
		$c_{021}$ & $\im \frac{(-1)^n}{4} e^{\im \psi}$ & $-\im \frac{(-1)^n}{4} e^{\im \psi}$ & $\im \frac{(-1)^n}{4} e^{\im \psi}$ & $-\im \frac{(-1)^n}{4} e^{\im \psi}$ \\[3pt]
		$c_{012}$ & $\frac{(-1)^n}{4} e^{\im (\psi+\varphi)}$ & $\frac{(-1)^n}{4} e^{\im (\psi+\varphi)}$ & $-\frac{(-1)^n}{4} e^{\im (\psi+\varphi)}$ & $-\frac{(-1)^n}{4} e^{\im (\psi+\varphi)}$ \\[3pt]
		\hline
		\hline
		$G$ &  \multicolumn{4}{c}{$\frac{\pi}{2} \left( 2m+1\right)$}\\[3pt]
		$\theta$ & $\frac{\pi}{4} \left( 2(4\tilde{n}+0)+1 \right)$ & $\frac{\pi}{4} \left( 2(4\tilde{n}+1)+1  \right)$ & $\frac{\pi}{4} \left( 2(4\tilde{n}+2)+1  \right)$ & $\frac{\pi}{4} \left( 2(4\tilde{n}+3)+1  \right)$ \\[3pt]
		\hline
		$c_{300}$ & $\im \frac{(-1)^m\sqrt{3}}{4} e^{-\im (2\psi+\varphi)}$ & $\im \frac{(-1)^m\sqrt{3}}{4} e^{-\im (2\psi+\varphi)}$ & $-\im \frac{(-1)^m\sqrt{3}}{4} e^{-\im (2\psi+\varphi)}$ & $-\im \frac{(-1)^m\sqrt{3}}{4} e^{-\im (2\psi+\varphi)}$ \\[3pt]
		$c_{003}$ & $\im \frac{(-1)^m\sqrt{3}}{4} e^{-\im (\psi+2\varphi)}$ & $-\im \frac{(-1)^m\sqrt{3}}{4} e^{-\im (\psi+2\varphi)}$ & $-\im \frac{(-1)^m\sqrt{3}}{4} e^{-\im (\psi+2\varphi)}$ & $\im \frac{(-1)^m\sqrt{3}}{4} e^{-\im (\psi+2\varphi)}$ \\[3pt]
		$c_{201}$ & $-\im \frac{(-1)^m}{4} e^{-\im \psi}$ & $\im \frac{(-1)^m}{4} e^{-\im \psi}$ & $\im \frac{(-1)^m}{4} e^{-\im \psi}$ & $-\im \frac{(-1)^m}{4} e^{-\im \psi}$ \\[3pt]
		$c_{120}$ & $\im \frac{(-1)^m}{2} e^{-\im \varphi}$ & $\im \frac{(-1)^m}{2} e^{-\im \varphi}$ & $-\im \frac{(-1)^m}{2} e^{-\im \varphi}$ & $-\im \frac{(-1)^m}{2} e^{-\im \varphi}$ \\[3pt]
		$c_{021}$ & $\im \frac{(-1)^m}{2} e^{\im \psi}$ & $-\im \frac{(-1)^m}{2} e^{\im \psi}$ & $-\im \frac{(-1)^m}{2} e^{\im \psi}$ & $\im \frac{(-1)^m}{2} e^{\im \psi}$ \\[3pt]
		$c_{102}$ & $-\im \frac{(-1)^m}{4} e^{\im \varphi}$ & $-\im \frac{(-1)^m}{4} e^{\im \varphi}$ & $\im \frac{(-1)^m}{4} e^{\im \varphi}$ & $\im \frac{(-1)^m}{4} e^{\im \varphi}$ \\[3pt]
		\hline
		\hline
		$G$ & $\pi \left( 2m+1-\frac{1}{3}\right)$ & $\pi \left( 2m+1+\frac{1}{3}\right)$ & $\pi \left( 2m+1-\frac{1}{3}\right)$ & $\pi \left( 2m+1+\frac{1}{3}\right)$ \\[3pt]
		$\theta$ & \multicolumn{2}{c|}{$\pi n +  \arctan\left( \sqrt{2}\right)$} & \multicolumn{2}{c}{$\pi n -  \arctan\left( \sqrt{2}\right)$}\\[3pt]
		\hline
		$c_{300}$ & $\im \frac{(-1)^n\sqrt{3}}{4} e^{-\im (2\psi+\varphi)}$ & $-\im \frac{(-1)^n\sqrt{3}}{4} e^{-\im (2\psi+\varphi)}$ & $-\im \frac{(-1)^n\sqrt{3}}{4} e^{-\im (2\psi+\varphi)}$ & $\im \frac{(-1)^n\sqrt{3}}{4} e^{-\im (2\psi+\varphi)}$ \\[3pt]
		$c_{030}$ &  $\frac{\sqrt{3}}{4} e^{\im (\psi-\varphi)}$ &  $\frac{\sqrt{3}}{4} e^{\im (\psi-\varphi)}$ &  $-\frac{\sqrt{3}}{4} e^{\im (\psi-\varphi)}$ &  $-\frac{\sqrt{3}}{4} e^{\im (\psi-\varphi)}$ \\[3pt]
		$c_{210}$ & $\frac{1}{4} e^{-\im(\psi+\varphi)}$ & $\frac{1}{4} e^{-\im(\psi+\varphi)}$ & $-\frac{1}{4} e^{-\im(\psi+\varphi)}$ & $-\frac{1}{4} e^{-\im(\psi+\varphi)}$ \\[3pt]
		$c_{120}$ & $\im \frac{(-1)^n}{4} e^{-\im \varphi}$ & $-\im \frac{(-1)^n}{4} e^{-\im \varphi}$ & $-\im \frac{(-1)^n}{4} e^{-\im \varphi}$ & $\im \frac{(-1)^n}{4} e^{-\im \varphi}$ \\[3pt]
		$c_{102}$ & $-\im \frac{(-1)^n}{2} e^{\im \varphi}$ & $\im \frac{(-1)^n}{2} e^{\im \varphi}$ & $\im \frac{(-1)^n}{2} e^{\im \varphi}$ & $-\im \frac{(-1)^n}{2} e^{\im \varphi}$ \\[3pt]
		$c_{012}$ & $\frac{1}{2} e^{\im (\psi+\varphi)}$ & $\frac{1}{2} e^{\im (\psi+\varphi)}$ & $-\frac{1}{2} e^{\im (\psi+\varphi)}$ & $\frac{1}{2} e^{\im (\psi+\varphi)}$\\[3pt]
		\end{tabular}
		\caption{Table of coefficients for equation (\ref{eq:ent31}). Coefficients, which are not listed, are equal to 0.}
		\label{tab:ent31}
\end{table*}

\begin{table*}
	\centering
	\footnotesize
		\begin{tabular}[b]{c|c|c}
		$G$ & \multicolumn{2}{c}{$2 \pi m + 2 \arctan\left( \sqrt{5 + 2 \sqrt{3}}\right)$}  \\[3pt]
		$\theta$ & $\pi n + \arctan\left( \frac{1}{2} (1 - \sqrt{3})\right)$ & $\pi n - \arctan\left( \frac{1}{2} (1 - \sqrt{3})\right)$ \\[3pt]
		\hline
		$c_{300}$ & $\im \frac{(-1)^n\sqrt{2}}{3} e^{-\im(2\psi+\varphi)}$ & $-\im \frac{(-1)^n\sqrt{2}}{3} e^{-\im(2\psi+\varphi)}$ \\[3pt]
		$c_{030}$ & $- \frac{1}{3\sqrt{2}} e^{\im (\psi-\varphi)}$ & $\frac{1}{3\sqrt{2}} e^{\im (\psi-\varphi)}$ \\[3pt]
		$c_{003}$ & $\im \frac{(-1)^n}{3\sqrt{2}} e^{\psi+2\varphi}$ & $\im \frac{(-1)^n}{3\sqrt{2}} e^{\psi+2\varphi}$ \\[3pt]
		$c_{120}$ & $\im \frac{(-1)^n}{\sqrt{6}} e^{-\im \varphi}$ & $-\im \frac{(-1)^n}{\sqrt{6}} e^{-\im \varphi}$ \\[3pt]
		$c_{021}$ & $\im \frac{(-1)^n}{\sqrt{6}} e^{\im \psi}$ & $\im \frac{(-1)^n}{\sqrt{6}} e^{\im \psi}$ \\[3pt]
		$c_{102}$ & $-\im \frac{(-1)^n}{\sqrt{6}} e^{\im \varphi}$ & $\im \frac{(-1)^n}{\sqrt{6}} e^{\im \varphi}$ \\[3pt]
		$c_{012}$ & $-\frac{(-1)^n}{\sqrt{6}} e^{\im (\psi + \varphi)}$ & $\frac{(-1)^n}{\sqrt{6}} e^{\im (\psi + \varphi)}$ \\[3pt]
		\hline
		\hline
		$G$ & \multicolumn{2}{c}{$2 \pi m - 2 \arctan\left( \sqrt{5 + 2 \sqrt{3}}\right)$} \\[3pt]
		$\theta$ & $\pi n + \arctan\left( \frac{1}{2} (1 - \sqrt{3})\right)$ & $\pi n - \arctan\left( \frac{1}{2} (1 - \sqrt{3})\right)$  \\[3pt]
		\hline
		$c_{300}$ & $-\im \frac{(-1)^n\sqrt{2}}{3} e^{-\im(2\psi+\varphi)}$ & $\im \frac{(-1)^n\sqrt{2}}{3} e^{-\im(2\psi+\varphi)}$ \\[3pt]
		$c_{030}$ & $- \frac{1}{3\sqrt{2}} e^{\im (\psi-\varphi)}$ & $\frac{1}{3\sqrt{2}} e^{\im (\psi-\varphi)}$ \\[3pt]
		$c_{003}$ & $-\im \frac{(-1)^n}{3\sqrt{2}} e^{\psi+2\varphi}$ & $-\im \frac{(-1)^n}{3\sqrt{2}} e^{\psi+2\varphi}$ \\[3pt]
		$c_{120}$ & $-\im \frac{(-1)^n}{\sqrt{6}} e^{-\im \varphi}$ & $\im \frac{(-1)^n}{\sqrt{6}} e^{-\im \varphi}$ \\[3pt]
		$c_{021}$ & $-\im \frac{(-1)^n}{\sqrt{6}} e^{\im \psi}$ & $-\im \frac{(-1)^n}{\sqrt{6}} e^{\im \psi}$ \\[3pt]
		$c_{102}$ & $\im \frac{(-1)^n}{\sqrt{6}} e^{\im \varphi}$ & $-\im \frac{(-1)^n}{\sqrt{6}} e^{\im \varphi}$ \\[3pt]
		$c_{012}$ & $-\frac{(-1)^n}{\sqrt{6}} e^{\im (\psi + \varphi)}$ & $\frac{(-1)^n}{\sqrt{6}} e^{\im (\psi + \varphi)}$ \\[3pt]
		\hline
		\hline
		$G$ & \multicolumn{2}{c}{$2 \pi m + 2 \arctan\left( \sqrt{5 - 2 \sqrt{3}}\right)$} \\[3pt]
		$\theta$ & $\pi n + \arctan\left( \frac{1}{2} (1 + \sqrt{3})\right)$ & $\pi n - \arctan\left( \frac{1}{2} (1 + \sqrt{3})\right)$\\[3pt]
		\hline
		$c_{300}$ & $\im \frac{(-1)^n\sqrt{2}}{3} e^{-\im(2\psi+\varphi)}$ & $-\im \frac{(-1)^n\sqrt{2}}{3} e^{-\im(2\psi+\varphi)}$ \\[3pt]
		$c_{030}$ & $\frac{1}{3\sqrt{2}} e^{\im (\psi-\varphi)}$ & $-\frac{1}{3\sqrt{2}} e^{\im (\psi-\varphi)}$ \\[3pt]
		$c_{003}$ & $\im \frac{(-1)^n}{3\sqrt{2}} e^{\psi+2\varphi}$ & $\im \frac{(-1)^n}{3\sqrt{2}} e^{\psi+2\varphi}$ \\[3pt]
		$c_{120}$ & $\im \frac{(-1)^n}{\sqrt{6}} e^{-\im \varphi}$ & $-\im \frac{(-1)^n}{\sqrt{6}} e^{-\im \varphi}$ \\[3pt]
		$c_{021}$ & $\im \frac{(-1)^n}{\sqrt{6}} e^{\im \psi}$ & $\im \frac{(-1)^n}{\sqrt{6}} e^{\im \psi}$ \\[3pt]
		$c_{102}$ & $-\im \frac{(-1)^n}{\sqrt{6}} e^{\im \varphi}$ & $\im \frac{(-1)^n}{\sqrt{6}} e^{\im \varphi}$ \\[3pt]
		$c_{012}$ & $\frac{(-1)^n}{\sqrt{6}} e^{\im (\psi + \varphi)}$ & $-\frac{(-1)^n}{\sqrt{6}} e^{\im (\psi + \varphi)}$ \\[3pt]
		\hline
		\hline
		$G$ & \multicolumn{2}{c}{$2 \pi m - 2 \arctan\left( \sqrt{5 - 2 \sqrt{3}}\right)$} \\[3pt]
		$\theta$ & $\pi n + \arctan\left( \frac{1}{2} (1 + \sqrt{3})\right)$ & $\pi n - \arctan\left( \frac{1}{2} (1 + \sqrt{3})\right)$  \\[3pt]
		\hline
		$c_{300}$ & $-\im \frac{(-1)^n\sqrt{2}}{3} e^{-\im(2\psi+\varphi)}$ & $\im \frac{(-1)^n\sqrt{2}}{3} e^{-\im(2\psi+\varphi)}$ \\[3pt]
		$c_{030}$ & $\frac{1}{3\sqrt{2}} e^{\im (\psi-\varphi)}$ & $-\frac{1}{3\sqrt{2}} e^{\im (\psi-\varphi)}$  \\[3pt]
		$c_{003}$ & $-\im \frac{(-1)^n}{3\sqrt{2}} e^{\psi+2\varphi}$ & $-\im \frac{(-1)^n}{3\sqrt{2}} e^{\psi+2\varphi}$ \\[3pt]
		$c_{120}$ & $-\im \frac{(-1)^n}{\sqrt{6}} e^{-\im \varphi}$ & $\im \frac{(-1)^n}{\sqrt{6}} e^{-\im \varphi}$ \\[3pt]
		$c_{021}$ & $-\im \frac{(-1)^n}{\sqrt{6}} e^{\im \psi}$ & $-\im \frac{(-1)^n}{\sqrt{6}} e^{\im \psi}$ \\[3pt]
		$c_{102}$ & $\im \frac{(-1)^n}{\sqrt{6}} e^{\im \varphi}$ & $\im \frac{(-1)^n}{\sqrt{6}} e^{\im \varphi}$ \\[3pt]
		$c_{012}$ & $\frac{(-1)^n}{\sqrt{6}} e^{\im (\psi + \varphi)}$ & $-\frac{(-1)^n}{\sqrt{6}} e^{\im (\psi + \varphi)}$ \\[3pt]
		\end{tabular}
		\caption{Table of coefficients for equation (\ref{eq:ent32}). Coefficients, which are not listed, are equal to 0.}
		\label{tab:ent32-1}
\end{table*}

\begin{table*}
	\centering
	\footnotesize
		\begin{tabular}[b]{c|c|c|c|c}
		$G$ & \multicolumn{4}{c}{$2 \pi m + \arctan\left( \sqrt{2 + \sqrt{3}} \right)$} \\[3pt]
		$\theta$ & $\frac{\pi}{4} \left( 2(4\tilde{n}+0)+1 \right)$ & $\frac{\pi}{4} \left( 2(4\tilde{n}+1)+1  \right)$ & $\frac{\pi}{4} \left( 2(4\tilde{n}+2)+1  \right)$ & $\frac{\pi}{4} \left( 2(4\tilde{n}+3)+1  \right)$\\[3pt]
		\hline
		$c_{300}$ & $\im \frac{1}{3\sqrt{2}} e^{-\im(2\psi+\varphi)}$ & $\im \frac{1}{3\sqrt{2}} e^{-\im(2\psi+\varphi)}$ & $-\im \frac{1}{3\sqrt{2}} e^{-\im(2\psi+\varphi)}$ & $-\im \frac{1}{3\sqrt{2}} e^{-\im(2\psi+\varphi)}$ \\[3pt]
		$c_{030}$ & $\frac{\sqrt{2}}{3} e^{\im (\psi-\varphi)}$ & $-\frac{\sqrt{2}}{3} e^{\im (\psi-\varphi)}$ & $\frac{\sqrt{2}}{3} e^{\im (\psi-\varphi)}$ & $-\frac{\sqrt{2}}{3} e^{\im (\psi-\varphi)}$ \\[3pt]
		$c_{003}$ & $\im \frac{1}{3\sqrt{2}} e^{\psi+2\varphi}$ & $-\im \frac{1}{3\sqrt{2}} e^{\psi+2\varphi}$ & $-\im \frac{1}{3\sqrt{2}} e^{\psi+2\varphi}$ & $\im \frac{1}{3\sqrt{2}} e^{\psi+2\varphi}$ \\[3pt]
		$c_{210}$ & $\frac{1}{\sqrt{6}} e^{-\im (\psi + \varphi)}$ & $-\frac{1}{\sqrt{6}} e^{-\im (\psi + \varphi)}$ & $\frac{1}{\sqrt{6}} e^{-\im (\psi + \varphi)}$ & $-\frac{1}{\sqrt{6}} e^{-\im (\psi + \varphi)}$ \\[3pt]
		$c_{201}$ & $-\im \frac{1}{\sqrt{6}} e^{-\im \psi}$ & $\im \frac{1}{\sqrt{6}} e^{-\im \psi}$ & $\im \frac{1}{\sqrt{6}} e^{-\im \psi}$ & $-\im \frac{1}{\sqrt{6}} e^{-\im \psi}$ \\[3pt]
		$c_{102}$ & $-\im \frac{1}{\sqrt{6}} e^{\im \varphi}$ & $-\im \frac{1}{\sqrt{6}} e^{\im \varphi}$ & $\im \frac{1}{\sqrt{6}} e^{\im \varphi}$ & $\im \frac{1}{\sqrt{6}} e^{\im \varphi}$ \\[3pt]
		$c_{012}$ & $\frac{1}{\sqrt{6}} e^{\im (\psi + \varphi)}$ & $-\frac{1}{\sqrt{6}} e^{\im (\psi + \varphi)}$ & $\frac{1}{\sqrt{6}} e^{\im (\psi + \varphi)}$ & $-\frac{1}{\sqrt{6}} e^{\im (\psi + \varphi)}$ \\[3pt]
		\hline
		\hline
		$G$ & \multicolumn{4}{c}{$2 \pi m - \arctan\left( \sqrt{2 + \sqrt{3}} \right)$} \\[3pt]
		$\theta$ & $\frac{\pi}{4} \left( 2(4\tilde{n}+0)+1 \right)$ & $\frac{\pi}{4} \left( 2(4\tilde{n}+1)+1  \right)$ & $\frac{\pi}{4} \left( 2(4\tilde{n}+2)+1  \right)$ & $\frac{\pi}{4} \left( 2(4\tilde{n}+3)+1  \right)$\\[3pt]
		\hline
		$c_{300}$ & $-\im \frac{1}{3\sqrt{2}} e^{-\im(2\psi+\varphi)}$ & $-\im \frac{1}{3\sqrt{2}} e^{-\im(2\psi+\varphi)}$ & $\im \frac{1}{3\sqrt{2}} e^{-\im(2\psi+\varphi)}$ & $\im \frac{1}{3\sqrt{2}} e^{-\im(2\psi+\varphi)}$ \\[3pt]
		$c_{030}$ & $\frac{\sqrt{2}}{3} e^{\im (\psi-\varphi)}$ & $-\frac{\sqrt{2}}{3} e^{\im (\psi-\varphi)}$ & $\frac{\sqrt{2}}{3} e^{\im (\psi-\varphi)}$ & $-\frac{\sqrt{2}}{3} e^{\im (\psi-\varphi)}$ \\[3pt]
		$c_{003}$ & $-\im \frac{1}{3\sqrt{2}} e^{\psi+2\varphi}$ & $\im \frac{1}{3\sqrt{2}} e^{\psi+2\varphi}$ & $\im \frac{1}{3\sqrt{2}} e^{\psi+2\varphi}$ & $-\im \frac{1}{3\sqrt{2}} e^{\psi+2\varphi}$ \\[3pt]
		$c_{210}$ & $\frac{1}{\sqrt{6}} e^{-\im (\psi + \varphi)}$ & $-\frac{1}{\sqrt{6}} e^{-\im (\psi + \varphi)}$ & $\frac{1}{\sqrt{6}} e^{-\im (\psi + \varphi)}$ & $-\frac{1}{\sqrt{6}} e^{-\im (\psi + \varphi)}$ \\[3pt]
		$c_{201}$ & $\im \frac{1}{\sqrt{6}} e^{-\im \psi}$ & $-\im \frac{1}{\sqrt{6}} e^{-\im \psi}$ & $-\im \frac{1}{\sqrt{6}} e^{-\im \psi}$ & $\im \frac{1}{\sqrt{6}} e^{-\im \psi}$ \\[3pt]
		$c_{102}$ & $\im \frac{1}{\sqrt{6}} e^{\im \varphi}$ & $\im \frac{1}{\sqrt{6}} e^{\im \varphi}$ & $-\im \frac{1}{\sqrt{6}} e^{\im \varphi}$ & $-\im \frac{1}{\sqrt{6}} e^{\im \varphi}$ \\[3pt]
		$c_{012}$ & $\frac{1}{\sqrt{6}} e^{\im (\psi + \varphi)}$ & $-\frac{1}{\sqrt{6}} e^{\im (\psi + \varphi)}$ & $\frac{1}{\sqrt{6}} e^{\im (\psi + \varphi)}$ & $-\frac{1}{\sqrt{6}} e^{\im (\psi + \varphi)}$ \\[3pt]
		\hline
		\hline
		$G$ & \multicolumn{4}{c}{$2 \pi m + \arctan\left( \sqrt{2 - \sqrt{3}} \right)$} \\[3pt]
		$\theta$ & $\frac{\pi}{4} \left( 2(4\tilde{n}+0)+1 \right)$ & $\frac{\pi}{4} \left( 2(4\tilde{n}+1)+1  \right)$ & $\frac{\pi}{4} \left( 2(4\tilde{n}+2)+1  \right)$ & $\frac{\pi}{4} \left( 2(4\tilde{n}+3)+1  \right)$\\[3pt]
		\hline
		$c_{300}$ & $\im \frac{1}{3\sqrt{2}} e^{-\im(2\psi+\varphi)}$ & $\im \frac{1}{3\sqrt{2}} e^{-\im(2\psi+\varphi)}$ & $-\im \frac{1}{3\sqrt{2}} e^{-\im(2\psi+\varphi)}$ & $-\im \frac{1}{3\sqrt{2}} e^{-\im(2\psi+\varphi)}$ \\[3pt]
		$c_{030}$ & $-\frac{\sqrt{2}}{3} e^{\im (\psi-\varphi)}$ & $\frac{\sqrt{2}}{3} e^{\im (\psi-\varphi)}$ & $-\frac{\sqrt{2}}{3} e^{\im (\psi-\varphi)}$ & $\frac{\sqrt{2}}{3} e^{\im (\psi-\varphi)}$ \\[3pt]
		$c_{003}$ & $\im \frac{1}{3\sqrt{2}} e^{\psi+2\varphi}$ & $-\im \frac{1}{3\sqrt{2}} e^{\psi+2\varphi}$ & $-\im \frac{1}{3\sqrt{2}} e^{\psi+2\varphi}$ & $\im \frac{1}{3\sqrt{2}} e^{\psi+2\varphi}$ \\[3pt]
		$c_{210}$ & $-\frac{1}{\sqrt{6}} e^{-\im (\psi + \varphi)}$ & $\frac{1}{\sqrt{6}} e^{-\im (\psi + \varphi)}$ & $-\frac{1}{\sqrt{6}} e^{-\im (\psi + \varphi)}$ & $\frac{1}{\sqrt{6}} e^{-\im (\psi + \varphi)}$ \\[3pt]
		$c_{201}$ & $-\im \frac{1}{\sqrt{6}} e^{-\im \psi}$ & $\im \frac{1}{\sqrt{6}} e^{-\im \psi}$ & $\im \frac{1}{\sqrt{6}} e^{-\im \psi}$ & $-\im \frac{1}{\sqrt{6}} e^{-\im \psi}$ \\[3pt]
		$c_{102}$ & $-\im \frac{1}{\sqrt{6}} e^{\im \varphi}$ & $-\im \frac{1}{\sqrt{6}} e^{\im \varphi}$ & $\im \frac{1}{\sqrt{6}} e^{\im \varphi}$ & $\im \frac{1}{\sqrt{6}} e^{\im \varphi}$ \\[3pt]
		$c_{012}$ & $-\frac{1}{\sqrt{6}} e^{\im (\psi + \varphi)}$ & $\frac{1}{\sqrt{6}} e^{\im (\psi + \varphi)}$ & $-\frac{1}{\sqrt{6}} e^{\im (\psi + \varphi)}$ & $\frac{1}{\sqrt{6}} e^{\im (\psi + \varphi)}$ \\[3pt]
		\hline
		\hline
		$G$ & \multicolumn{4}{c}{$2 \pi m - \arctan\left( \sqrt{2 - \sqrt{3}} \right)$} \\[3pt]
		$\theta$ & $\frac{\pi}{4} \left( 2(4\tilde{n}+0)+1 \right)$ & $\frac{\pi}{4} \left( 2(4\tilde{n}+1)+1  \right)$ & $\frac{\pi}{4} \left( 2(4\tilde{n}+2)+1  \right)$ & $\frac{\pi}{4} \left( 2(4\tilde{n}+3)+1  \right)$\\[3pt]
		\hline
		$c_{300}$ & $-\im \frac{1}{3\sqrt{2}} e^{-\im(2\psi+\varphi)}$ & $-\im \frac{1}{3\sqrt{2}} e^{-\im(2\psi+\varphi)}$ & $\im \frac{1}{3\sqrt{2}} e^{-\im(2\psi+\varphi)}$ & $\im \frac{1}{3\sqrt{2}} e^{-\im(2\psi+\varphi)}$ \\[3pt]
		$c_{030}$ & $-\frac{\sqrt{2}}{3} e^{\im (\psi-\varphi)}$ & $\frac{\sqrt{2}}{3} e^{\im (\psi-\varphi)}$ & $-\frac{\sqrt{2}}{3} e^{\im (\psi-\varphi)}$ & $\frac{\sqrt{2}}{3} e^{\im (\psi-\varphi)}$ \\[3pt]
		$c_{003}$ & $-\im \frac{1}{3\sqrt{2}} e^{\psi+2\varphi}$ & $\im \frac{1}{3\sqrt{2}} e^{\psi+2\varphi}$ & $\im \frac{1}{3\sqrt{2}} e^{\psi+2\varphi}$ & $-\im \frac{1}{3\sqrt{2}} e^{\psi+2\varphi}$ \\[3pt]
		$c_{210}$ & $-\frac{1}{\sqrt{6}} e^{-\im (\psi + \varphi)}$ & $\frac{1}{\sqrt{6}} e^{-\im (\psi + \varphi)}$ & $-\frac{1}{\sqrt{6}} e^{-\im (\psi + \varphi)}$ & $\frac{1}{\sqrt{6}} e^{-\im (\psi + \varphi)}$ \\[3pt]
		$c_{201}$ & $\im \frac{1}{\sqrt{6}} e^{-\im \psi}$ & $-\im \frac{1}{\sqrt{6}} e^{-\im \psi}$ & $-\im \frac{1}{\sqrt{6}} e^{-\im \psi}$ & $\im \frac{1}{\sqrt{6}} e^{-\im \psi}$ \\[3pt]
		$c_{102}$ & $\im \frac{1}{\sqrt{6}} e^{\im \varphi}$ & $\im \frac{1}{\sqrt{6}} e^{\im \varphi}$ & $-\im \frac{1}{\sqrt{6}} e^{\im \varphi}$ & $-\im \frac{1}{\sqrt{6}} e^{\im \varphi}$ \\[3pt]
		$c_{012}$ & $-\frac{1}{\sqrt{6}} e^{\im (\psi + \varphi)}$ & $\frac{1}{\sqrt{6}} e^{\im (\psi + \varphi)}$ & $-\frac{1}{\sqrt{6}} e^{\im (\psi + \varphi)}$ & $\frac{1}{\sqrt{6}} e^{\im (\psi + \varphi)}$ \\[3pt]
		\end{tabular}
		\caption{Table of coefficients for equation (\ref{eq:ent32}). Coefficients, which are not listed, are equal to 0.}
		\label{tab:ent32-2}
\end{table*}

\begin{table*}
	\centering
	\footnotesize
		\begin{tabular}[b]{c|c|c}
		$G$ & \multicolumn{2}{c}{$2 \pi m + 2 \arctan\left( \sqrt{5 + 2 \sqrt{3}}\right)$}  \\[3pt]
		$\theta$ & $\pi n + \arctan\left( 1 + \sqrt{3}\right)$ & $\pi n - \arctan\left( 1 + \sqrt{3}\right)$ \\[3pt]
		\hline
		$c_{300}$ & $\im \frac{(-1)^n}{3\sqrt{2}} e^{-\im(2\psi+\varphi)}$ & $-\im \frac{(-1)^n}{3\sqrt{2}} e^{-\im(2\psi+\varphi)}$\\[3pt]
		$c_{030}$ & $\frac{1}{3\sqrt{2}} e^{\im (\psi-\varphi)}$ & $-\frac{1}{3\sqrt{2}} e^{\im (\psi-\varphi)}$ \\[3pt]
		$c_{003}$ & $-\im \frac{(-1)^n\sqrt{2}}{3} e^{\psi+2\varphi}$  & $-\im \frac{(-1)^n\sqrt{2}}{3} e^{\psi+2\varphi}$ \\[3pt]
		$c_{210}$ & $\frac{1}{\sqrt{6}} e^{-\im (\psi + \varphi)}$  & $-\frac{1}{\sqrt{6}} e^{-\im (\psi + \varphi)}$ \\[3pt]
		$c_{201}$ & $\im \frac{(-1)^n}{\sqrt{6}} e^{-\im \psi}$  & $\im \frac{(-1)^n}{\sqrt{6}} e^{-\im \psi}$ \\[3pt]
		$c_{120}$ & $\im \frac{(-1)^n}{\sqrt{6}} e^{-\im \varphi}$  & $-\im \frac{(-1)^n}{\sqrt{6}} e^{-\im \varphi}$ \\[3pt]
		$c_{021}$ & $-\im \frac{(-1)^n}{\sqrt{6}} e^{\im \psi}$  & $-\im \frac{(-1)^n}{\sqrt{6}} e^{\im \psi}$ \\[3pt]
		\hline
		\hline
		$G$ & \multicolumn{2}{c}{$2 \pi m - 2 \arctan\left( \sqrt{5 + 2 \sqrt{3}}\right)$}  \\[3pt]
		$\theta$ & $\pi n + \arctan\left( 1 + \sqrt{3}\right)$ & $\pi n - \arctan\left( 1 + \sqrt{3}\right)$ \\[3pt]
		\hline
		$c_{300}$ & $-\im \frac{(-1)^n}{3\sqrt{2}} e^{-\im(2\psi+\varphi)}$ & $\im \frac{(-1)^n}{3\sqrt{2}} e^{-\im(2\psi+\varphi)}$\\[3pt]
		$c_{030}$ & $\frac{1}{3\sqrt{2}} e^{\im (\psi-\varphi)}$ & $-\frac{1}{3\sqrt{2}} e^{\im (\psi-\varphi)}$ \\[3pt]
		$c_{003}$ & $\im \frac{(-1)^n\sqrt{2}}{3} e^{\psi+2\varphi}$  & $\im \frac{(-1)^n\sqrt{2}}{3} e^{\psi+2\varphi}$ \\[3pt]
		$c_{210}$ & $\frac{1}{\sqrt{6}} e^{-\im (\psi + \varphi)}$  & $-\frac{1}{\sqrt{6}} e^{-\im (\psi + \varphi)}$ \\[3pt]
		$c_{201}$ & $-\im \frac{(-1)^n}{\sqrt{6}} e^{-\im \psi}$  & $-\im \frac{(-1)^n}{\sqrt{6}} e^{-\im \psi}$ \\[3pt]
		$c_{120}$ & $-\im \frac{(-1)^n}{\sqrt{6}} e^{-\im \varphi}$  & $\im \frac{(-1)^n}{\sqrt{6}} e^{-\im \varphi}$ \\[3pt]
		$c_{021}$ & $\im \frac{(-1)^n}{\sqrt{6}} e^{\im \psi}$  & $\im \frac{(-1)^n}{\sqrt{6}} e^{\im \psi}$ \\[3pt]
		\hline
		\hline
		$G$ & \multicolumn{2}{c}{$2 \pi m + 2 \arctan\left( \sqrt{5 - 2 \sqrt{3}}\right)$}  \\[3pt]
		$\theta$ & $\pi n + \arctan\left( 1 - \sqrt{3}\right)$ & $\pi n - \arctan\left( 1 - \sqrt{3}\right)$ \\[3pt]
		\hline
		$c_{300}$ & $-\im \frac{(-1)^n}{3\sqrt{2}} e^{-\im(2\psi+\varphi)}$ & $\im \frac{(-1)^n}{3\sqrt{2}} e^{-\im(2\psi+\varphi)}$\\[3pt]
		$c_{030}$ & $-\frac{1}{3\sqrt{2}} e^{\im (\psi-\varphi)}$ & $\frac{1}{3\sqrt{2}} e^{\im (\psi-\varphi)}$ \\[3pt]
		$c_{003}$ & $\im \frac{(-1)^n\sqrt{2}}{3} e^{\psi+2\varphi}$  & $\im \frac{(-1)^n\sqrt{2}}{3} e^{\psi+2\varphi}$ \\[3pt]
		$c_{210}$ & $-\frac{1}{\sqrt{6}} e^{-\im (\psi + \varphi)}$  & $\frac{1}{\sqrt{6}} e^{-\im (\psi + \varphi)}$ \\[3pt]
		$c_{201}$ & $-\im \frac{(-1)^n}{\sqrt{6}} e^{-\im \psi}$  & $-\im \frac{(-1)^n}{\sqrt{6}} e^{-\im \psi}$ \\[3pt]
		$c_{120}$ & $-\im \frac{(-1)^n}{\sqrt{6}} e^{-\im \varphi}$  & $\im \frac{(-1)^n}{\sqrt{6}} e^{-\im \varphi}$ \\[3pt]
		$c_{021}$ & $\im \frac{(-1)^n}{\sqrt{6}} e^{\im \psi}$  & $\im \frac{(-1)^n}{\sqrt{6}} e^{\im \psi}$ \\[3pt]
		\hline
		\hline
		$G$ & \multicolumn{2}{c}{$2 \pi m - 2 \arctan\left( \sqrt{5 - 2 \sqrt{3}}\right)$}  \\[3pt]
		$\theta$ & $\pi n + \arctan\left( 1 - \sqrt{3}\right)$ & $\pi n - \arctan\left( 1 - \sqrt{3}\right)$ \\[3pt]
		\hline
		$c_{300}$ & $\im \frac{(-1)^n}{3\sqrt{2}} e^{-\im(2\psi+\varphi)}$ & $-\im \frac{(-1)^n}{3\sqrt{2}} e^{-\im(2\psi+\varphi)}$\\[3pt]
		$c_{030}$ & $-\frac{1}{3\sqrt{2}} e^{\im (\psi-\varphi)}$ & $\frac{1}{3\sqrt{2}} e^{\im (\psi-\varphi)}$ \\[3pt]
		$c_{003}$ & $-\im \frac{(-1)^n\sqrt{2}}{3} e^{\psi+2\varphi}$  & $-\im \frac{(-1)^n\sqrt{2}}{3} e^{\psi+2\varphi}$ \\[3pt]
		$c_{210}$ & $-\frac{1}{\sqrt{6}} e^{-\im (\psi + \varphi)}$  & $\frac{1}{\sqrt{6}} e^{-\im (\psi + \varphi)}$ \\[3pt]
		$c_{201}$ & $\im \frac{(-1)^n}{\sqrt{6}} e^{-\im \psi}$  & $\im \frac{(-1)^n}{\sqrt{6}} e^{-\im \psi}$ \\[3pt]
		$c_{120}$ & $\im \frac{(-1)^n}{\sqrt{6}} e^{-\im \varphi}$  & $-\im \frac{(-1)^n}{\sqrt{6}} e^{-\im \varphi}$ \\[3pt]
		$c_{021}$ & $-\im \frac{(-1)^n}{\sqrt{6}} e^{\im \psi}$  & $-\im \frac{(-1)^n}{\sqrt{6}} e^{\im \psi}$ \\[3pt]
		\end{tabular}
		\caption{Table of coefficients for equation (\ref{eq:ent32}). Coefficients, which are not listed, are equal to 0.}
		\label{tab:ent32-3}
\end{table*}

\end{document}